# First implementation of full-workflow automation in radiotherapy: the All-in-One solution on rectal cancer

Lei Yu[*,1,2], Jun Zhao[*,1,2], Fan Xia[1,2], Zhiyuan Zhang[1,2], Yanfang Liu[3], Wei Zhang[3], Jingjie Zhou[3], Jiazhou Wang[#,1,2], Weigang Hu[#,1,2], Zhen Zhang[#,1,2]

[1] Department of Radiation Oncology, Fudan University Shanghai Cancer Center, Shanghai, China

[2] Department of Oncology, Shanghai Medical College, Fudan University, Shanghai, China

[3] Shanghai United Imaging Intelligence Inc., Shanghai, China

* These authors contributed equally.

# These authors are corresponding authors.

**Abstract:**

The aim of this work is to describe the technical characteristics of an AI-powered radiotherapy workflow that enables full-process automation (All-in-One), evaluate its performance implemented for on-couch initial treatment of rectal cancer, and provide insight into the behavior of full-workflow automation in the specialty of radiotherapy. The All-in-One workflow was developed based on a CT-integrated linear accelerator. It incorporates routine radiotherapy procedures from simulation, autosegmentation, autoplanning, image guidance, beam delivery, and in vivo quality assurance (QA) into one scheme, with critical decision points involved, while the patient is on the treatment couch during the whole process.

For the enrolled ten patients with rectal cancer, minor modifications of the autosegmented target volumes were required, and the Dice similarity coefficient and 95% Hausdorff distance before and after modifications were 0.892±0.061 and 18.2±13.0 mm, respectively. The autosegmented normal tissues and automatic plans were clinically acceptable without any modifications or reoptimization. The pretreatment IGRT corrections were within 2 mm in all directions, and the EPID-based in vivo QA showed a γ passing rate better than 97% (3%/3 mm/10% threshold). The duration of the whole process was 23.2±3.5 minutes, depending mostly on the time required for manual modification and plan evaluation.

The All-in-One workflow enables full automation of the entire radiotherapy process by seamlessly integrating multiple routine procedures. The one-stop solution shortens the time scale it takes to ready the first treatment from days to minutes, significantly improving the patient experience and the efficiency of the workflow, and shows potential to facilitate the clinical application of online adaptive replanning.

1. **Introduction**

Radiotherapy (RT) is performed in a multistep workflow with complicated physics and mathematics involved, relying mostly on human efforts, where various uncertainties may exist and affect the treatment efficacy. On the one hand, manual delineation of target volumes and normal tissues as well as treatment planning design are very time-consuming and may suffer from inter- and intraoperator variability, depending on the skill and expertise of the operator [1]. For patients who undergo radiation therapy, it usually takes several days up to a couple of weeks from computed tomography (CT) simulation to the first treatment. The longer the patient waits, the more likely his or her condition (tumor progression, weight, anatomic structure, etc.) is to differ from the planning CT, leading to the delivered dose deviating from the planned dose [2-3]. On the other hand, despite the extensive use of image-guided RT (IGRT) [4], intra- and/or interfractional variations in patient anatomy across the RT course frequently occur, which could be a source of deteriorated treatment accuracy, thus compromising tumor control [5-6]. Therefore, the conventional RT workflow needs to be improved to make the treatment more accurate and efficient.

Recent progress in artificial intelligence (AI) technology, especially deep learning coupled with increasing clinical data, is considered to provide promising solutions that could standardize and speed up RT procedures [7], e.g., automated segmentation [8] and automated treatment planning [9]. Various AI applications are now clinically available in commercial treatment systems, such as the atlas-based contouring tool and the knowledge-based planning (KBP) module. These AI solutions are able to achieve comparable performance with handwork in segmentation accuracy and plan quality, with only minor editing but marked efficiency improvement [10-11].

Recently, an AI-based RT workflow that integrates autosegmentation and autoplanning based on scripts in Pinnacle³ (Philips Medical Systems, Madison, WI) has been reported, showing the potential of automation for reducing workloads in clinical practice [12]. However, there is still a gap between partial automation and full-workflow automation. Fragmentary implementation of automation usually needs extra data processing and transmission, especially between different platforms, and results in a wide set of endpoints and complex (mostly repetitive) user interactions, holding it back in efficiency improvement.

In this contribution, we proposed a streamlined AI-driven online workflow for full-process automation of initial treatment based on a CT-integrated linear accelerator (linac). The so-called "All-in-One" workflow incorporates the multiple RT steps from simulation, autosegmentation, autoplanning, image guidance, beam delivery, and in vivo patient-specific quality assurance (QA) into one scheme. Critical decision points for the evaluation of contouring, planning, IGRT, and QA results are designed to secure the treatment and drive forward the automation process.

The new workflow has been implemented in practice for a cohort of 10 patients with rectal cancer. The entire process from simulation to the first delivery can be seamlessly performed while the patient is on the treatment couch. In this article, the technical characteristics of the All-in-One workflow will be outlined, and the patient data of clinical implementation will be presented, including the outcomes of autosegmentation and autoplanning, the results of pretreatment image guidance and in vivo QA, and the duration of each phase. Clinical gain and loss brought by the full-process automation will also be discussed.

## 2. Materials and methods

We developed the All-in-One workflow based on a recently commercialized CT-integrated linac, uRT-linac 506c (United Imaging Healthcare, UIH, Shanghai, China). Herein we will briefly describe the basic design of the treatment system, and illustrate the procedures for the configuration, implementation, and evaluation of the All-in-One workflow.

### 2.1 System design

The uRT-linac 506c platform combines a diagnostic-quality 16-slice helical CT scanner with a C-arm linac [13]. The on-board CT scanner has a bore diameter of 70 cm, attached coaxially behind the linac gantry with a longitudinal distance of 2100 mm between the treatment isocenter and CT origin. The hybrid system is incorporated with a proprietary control system consisting of a treatment planning and oncology information system (uRT-TPOIS) and a treatment delivery system with a shared patient database. The uRT-TPOIS offers embedded AI modules for autosegmentation and autoplanning. A multiresolution VB-Net convolutional neural network (CNN) is utilized for automated segmentation [14]. Customized models are supported based on the clinical patient database of local institutions. Automated planning is implemented by voxel-based optimization according to preset clinical goals combined with built-in control strategies (target homogeneity, conformity, cold/hot spot control, dose control of organ at risk, etc.).

In vivo dosimetry is available by measuring the exit dose from the patient using EPID. The measured transit dose distribution is compared with the TPS calculation based on patient anatomy using a Monte Carlo method, and the γ index is evaluated at the end of each delivered beam. Prior to clinical implementation, the EPID detector response was corrected and calibrated for absolute dosimetry, and the accuracy of the transit dose calculation was validated by phantom measurements [15].

**2.2 Treatment preparation**

Before the treatment, site-specific protocols for CT scanning, autosegmentation, autoplanning, and optimization objectives need to be configured, as described below.

- CT scanning protocol: Protocol of head/thorax/abdomen/pelvis is specified; thus, a proper scanning range (corresponding to the longitudinal coordinate of the couch) can be automatically determined.

- Autosegmentation protocol: Target volumes and organs-at-risk (OARs) that need to be autosegmented are defined. In particular, margin expansion between the clinical target volume (CTV) and planning target volume (PTV) also is included. When the CTV is manually revised, the PTV can be updated immediately.

- Autoplanning protocol: This protocol includes the delivery technique, plan normalization mode, beam angle configuration, and plan optimization hyperparameters.

- Optimization objective protocol: dose prescription on PTV and dose constraints for OARs are set.

In this study, the target and OAR segmentation models were trained and validated based on 195 rectal patients selected randomly from the clinical database of our

institution from 2016 to 2018.

**2.3 Workflow procedures**

Figure 1 illustrates the flow chart of the All-in-One workflow, and the detailed procedures are the following.

- Position the patient on the couch, and launch the All-in-One workflow at the treatment console by sending the patient to the CT scanner to start scanning. The CT series together with the setup coordinates are automatically sent to the uRT-TPOIS.

- In uRT-TPOIS, target volumes and OARs are autosegmented. The oncologist reviews the contouring results combined with clinical diagnosis and makes adjustments if necessary.

- A new plan with preset clinical goals is generated, where the isocenter is specified as the geometric center of the PTV. The corresponding CT-to-density table is automatically assigned and the couch structure is added by detecting its position on the image. The above plan parameters are reviewed by the physicist and modified if necessary.

- Auto-plan optimization is started. Meanwhile, the offset between the setup and isocenter is sent to the linac to allow couch motion. The patient is aligned to the treatment isocenter by automatic couch repositioning. The therapist enters the treatment room and marks the isocenter position on the patient surface for the sake of subsequent fractions.

- The auto-plan result is reviewed by the physicist and adjusted as needed, and

then evaluated and approved by the oncologist.

- The EPID transit dose of each beam is calculated and used for in vivo QA. The plan is finally approved and scheduled for treatment.

- Pretreatment image guidance using either MV orthogonal portal imaging or low-dose CT is carried out. The acquired image is registered to the planning CT and visually inspected by the therapist.

- Treatment beams are delivered while the transit dose is measured by EPID. An action limit of 90% is used for the γ analysis of in vivo QA (global normalization in absolute dose, 3% dose difference, 3 mm distance-to-agreement, and 10% threshold).

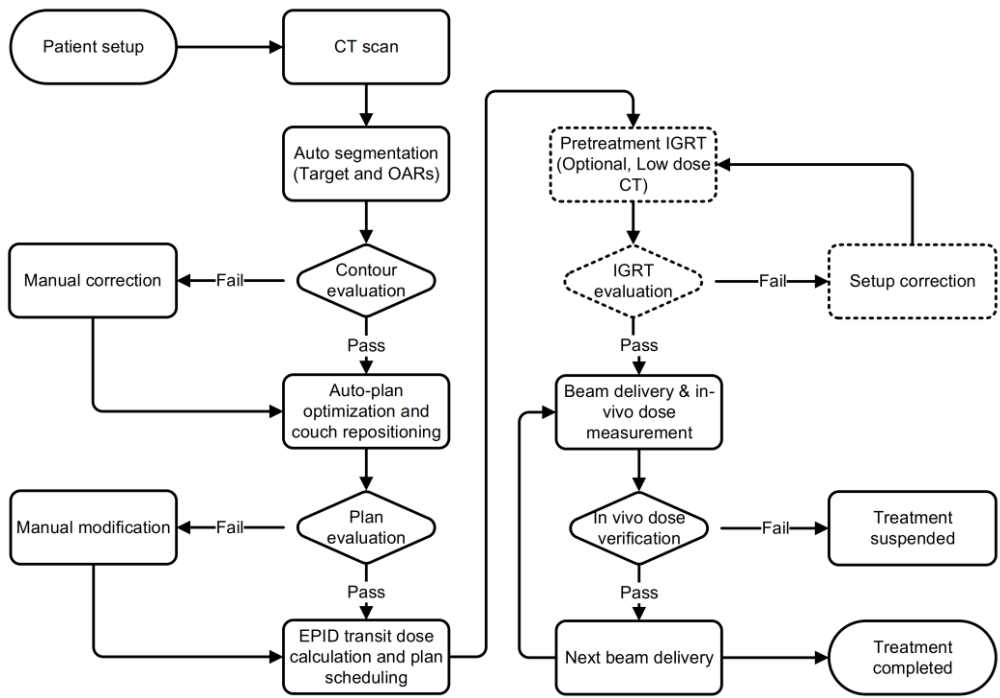

Fig. 1: Flow chart of the All-in-One solution

## 2.4 Clinical implementation

A cohort of 10 patients with rectal cancer who were going to receive neoadjuvant radiotherapy were enrolled for clinical implementation of the new workflow. The patients were prescribed either long-course radiotherapy (LCRT) at 50 Gy/5f (4 cases) or short-course radiotherapy (SCRT) at 25 Gy/5f (6 cases). Templates of the All-in-One protocols are listed in Table 1. The patients were positioned supinely with headrest and footrest for immobilization and were advised to remain stationary until the end of the treatment. Each patient was CT-simulated with a 3-mm slice thickness, and received a low-dose CT for pretreatment IGRT, and then was treated with one-arc VMAT. An experienced radiation oncologist, a medical physicist, and two therapists participated in the whole process.

Table 1: All-in-One protocols for the enrolled patients with rectal cancer

| All-in-One protocols | Details | |
|---|---|---|
| | 50 Gy in 25 fractions (LCRT) | 25 Gy in 5 fractions (SCRT) |
| CT scanning protocol | Pelvis protocol | |
| Autosegmentation protocol | Bladder, left femoral head, right femoral head, small bowel, colon, CTV | |
| | CTV-to-PTV margin: L-R: 0.7 cm, S-I: 0.5 cm, A-P: 0.6 cm | CTV-to-PTV margin: L-R: 0.5 cm, S-I: 0.5 cm, A-P: 0.5 cm |
| Autoplanning protocol | One-arc VMAC, 30 iterations | |
| | Plan normalization: 100% prescription dose covers 97% of PTV | |
| Optimization objective protocol | PTV: 50 Gy prescription, Dmax<52.5 Gy | PTV: 25 Gy prescription, Dmax<26.25 Gy |
| | Bladder: V45Gy<30% | Bladder: V20Gy<30% |
| | Left/right femoral head: Dmean<20 Gy | Left/right femoral head: Dmean<8 Gy |

**2.5 Performance evaluation**

The performance of the All-in-One solution was evaluated by the attendant staff in terms of segmentation accuracy, plan quality, pretreatment positioning stability, in vivo QA results, and total duration. For the autosegmented structures, the geometric difference before and after correction was quantitatively investigated by analyzing the Dice similarity coefficient (DSC) and 95% Hausdorff distance (HD_95, the 95th percentile of the distances to eliminate the impact of a small fraction of outliers).

The results of automated plans were evaluated using the following dose volume indices. For the PTV, relative volumes covered by 95%, 99%, and 105% of the prescribed dose ($V_{95\%}$, $V_{99\%}$, and $V_{105\%}$), conformity index (CI) and homogeneity index (HI) were evaluated, where the CI was defined as,

$$CI = \frac{(V_{PTV}^{RI})^2}{V_{RI} \times V_{PTV}}, \quad (1)$$

where $V_{RI}$, $V_{PTV}$, and $V_{PTV}^{RI}$ are the volume covered by the prescribed dose, the target volume, and the target volume covered by the prescribed dose, respectively. And,

$$HI = \frac{D_2 - D_{98}}{D_p}, \quad (2)$$

where $D_2$, $D_{98}$, and $D_p$ are the dose to 2% of the target volume, dose to 98% of the target volume, and the prescribed dose, respectively. For the OARs, the mean dose ($D_{mean}$), $V_{25Gy}$, $V_{35Gy}$, $V_{45Gy}$ of the bladder and the $D_{mean}$, $D_2$, $V_{30Gy}$, $V_{45Gy}$ of the bilateral femoral heads (FHs) were used, where $V_{XXGy}$ means the relative volume of the OAR covered by an isodose line of XX Gy. The $V_{30Gy}$, $V_{35Gy}$, $V_{45Gy}$ of the small bowel and the $V_{50Gy}$ of the colon were evaluated in absolute volume. For the patients with SCRT, dose-volume indices of the OARs were evaluated by converting the isodoses to the corresponding equivalent doses in 2 Gy/f ($EQD_2$) [16].

## 3. Results

Compared to more than 30 min by manual delineation, it took the oncologist 3 min to 8 min to modify the autosegmented CTV, and the autosegmented OARs (bladder, FHs, small bowel, colon) were checked by the oncologist in the entire scan area and clinically accepted without any modification. The calculated DSC and HD_95 metrics between the autosegmented and manually corrected CTVs were 0.892±0.061 and 18.2±13.0 mm (mean ± standard deviation), respectively. The results of automated planning in target coverage and OAR sparing were well within clinical criteria and were approved in situ without the need for reoptimization, as shown in Table 2. An example of the autosegmented structures with isodoses of autoplanning is illustrated in Fig. 2.

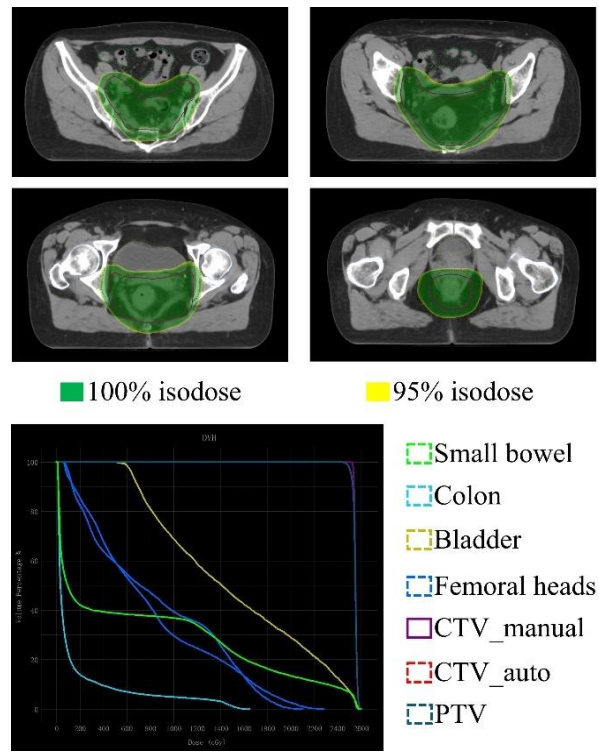

Fig. 2: An example of autosegmentation (shown in dashed line), manually corrected CTV (solid line), and the result of autoplanning (isodoses in filled area and DVH plot).

Table 2: Dose-volume indices of the autoplanning results for the enrolled patients with LCRT and SCRT. For the SCRT patients, the dose-volume indices of the OARs were evaluated in $EQD_2$.

| | LCRT: 50Gy/25f | | SCRT: 25Gy/5f | |
|---|---|---|---|---|
| DVH indices | All-in-One plan (mean ± standard deviation) | DVH indices | All-in-One plan (mean ± standard deviation) | |
| PTV_Dmax (Gy) | 52.36±0.19 | PTV_Dmax (Gy) | 26.14±0.10 |
| PTV_V95% (%) | 100 | PTV_V95% (%) | 100 |
| PTV_V99% (%) | 99.42±0.14 | PTV_V99% (%) | 99.37±0.12 |
| PTV_V105% (%) | 0.01±0.01 | PTV_V105% (%) | 0.01±0.01 |
| PTV_CI | 0.94±0.01 | PTV_CI | 0.95±0.01 |
| PTV_HI | 0.04±0.01 | PTV_HI | 0.03±0.01 |
| Bladder_Dmean (Gy) | 40.25±5.54 | Bladder_Dmean (Gy) | 14.76±0.91 |
| Bladder_V25Gy (%) | 97.40±4.26 | Bladder_V16.67Gy (%) * | 39.14±6.72 |
| Bladder_V35Gy (%) | 63.72±20.06 | Bladder_V23.33Gy (%) * | 15.85±8.71 |
| Bladder_V45Gy (%) | 41.47±29.71 | Bladder_V30Gy (%) * | 0 |
| LeftFH_Dmean (Gy) | 18.95±1.35 | LeftFH_Dmean (Gy) | 8.52±0.57 |
| LeftFH_D2 (Gy) | 41.93±3.92 | LeftFH_D2 (Gy) | 17.74±1.62 |
| LeftFH_V25Gy (%) | 26.14±6.53 | LeftFH_V12.07Gy (%) * | 24.49±7.25 |
| LeftFH_V40Gy (%) | 5.30±4.91 | LeftFH_V19.31Gy (%) * | 1.11±1.38 |
| RightFH_Dmean (Gy) | 21.68±2.62 | RightFH_Dmean (Gy) | 8.65±0.53 |
| RightFH_D2 (Gy) | 40.86±3.69 | RightFH_D2 (Gy) | 18.58±0.84 |
| RightFH_V25Gy (%) | 35.33±8.81 | RightFH_V12.07Gy (%) * | 30.83±4.11 |
| RightFH_V40Gy (%) | 4.28±4.27 | RightFH_V19.31Gy (%) * | 1.36±1.82 |
| Small bowel_D2 (Gy) | 45.66±5.44 | Small bowel_D2 (Gy) | 23.50±3.00 |
| Small bowel_V30Gy (ml) | 71.31±20.34 | Small bowel_V22.5Gy (ml) * | 39.03±24.16 |
| Small bowel_V35Gy (ml) | 40.18±16.10 | Small bowel_V26.25Gy (ml) * | 0 |
| Small bowel_V45Gy (ml) | 16.48±9.91 | Small bowel_V33.76Gy (ml) * | 0 |
| Colon_V50Gy (ml) | 0 | Colon_V33.33Gy (ml) * | 0 |
| Colon_D2 (Gy) | 26.46±9.96 | Colon_D2 (Gy) | 8.48±5.39 |

*α/β values for EQD2 calculation were taken averagely from Ref. [16]. Bladder: α/β=4. FH: α/β=0.8. Small bowel: α/β=7. Colon: α/β= 4.

Table 3 summarizes the treatment data of the enrolled patients, including the prescription, IGRT correction, average γ passing rate of in vivo QA, and duration of the first treatment. The IGRT couch correction prior to the first fraction was basically within 2 mm in all directions. The γ index of in vivo QA was checked every 30 degrees, and all the checkpoints suggested a passing rate better than 97% (3%/3 mm/10% threshold). The total duration from the beginning of simulation to the end of beam delivery was 23.2±3.5 minutes, and the time expended in each phase of the workflow is depicted in Fig. 3. It shows that fine tuning of the target contour and plan evaluation took up the majority of the overall time, which is expected to be accelerated by improving the segmentation model and process integration. Patient 8 spent more time in the phase of plan evaluation because the oncologist edited the target contour once again after the plan optimization was finished.

Table 3: Treatment data of the enrolled patients in All-in-One implementation

| Patient | Prescription | IGRT correction (VRT, LNG, LAT) prior to the first delivery | γ passing rate of in vivo QA in the first delivery | Total duration |
|---|---|---|---|---|
| 1 | 2Gy*25F | (0.2 mm, 0.5 mm, -0.3 mm) | 98.4 ± 0.8% | 23.0 min |
| 2 | 5Gy*5F | (0.7 mm, -0.2 mm, 1.5 mm) | 99.6 ± 0.5% | 26.4 min |
| 3 | 5Gy*5F | (0.8 mm, 0.4 mm, 0.9 mm) | 98.8 ± 1.0% | 22.5 min |
| 4 | 5Gy*5F | (0.2 mm, 0.2 mm, -0.7 mm) | 99.8 ± 0.2% | 21.8 min |
| 5 | 2Gy*25F | (0.4 mm, 1.0 mm, 0.3 mm) | 99.8 ± 0.2% | 21.7 min |
| 6 | 2Gy*25F | (0.7 mm, 1.5 mm, -0.5 mm) | 99.8 ± 0.3% | 27.3 min |
| 7 | 2Gy*25F | (0.3 mm, 0.9 mm, 1.0 mm) | 99.8 ± 0.2% | 17.8 min |
| 8 | 5Gy*5F | (0.9 mm, 1.8 mm, -0.1 mm) | 99.7 ± 0.4% | 30.0 min |
| 9 | 5Gy*5F | (-0.1 mm, 0.5 mm, -1.2 mm) | 99.5 ± 0.6% | 21.7 min |
| 10 | 5Gy*5F | (0.2 mm, 0.4 mm, 0.3 mm) | 99.9 ± 0.1% | 19.7 min |

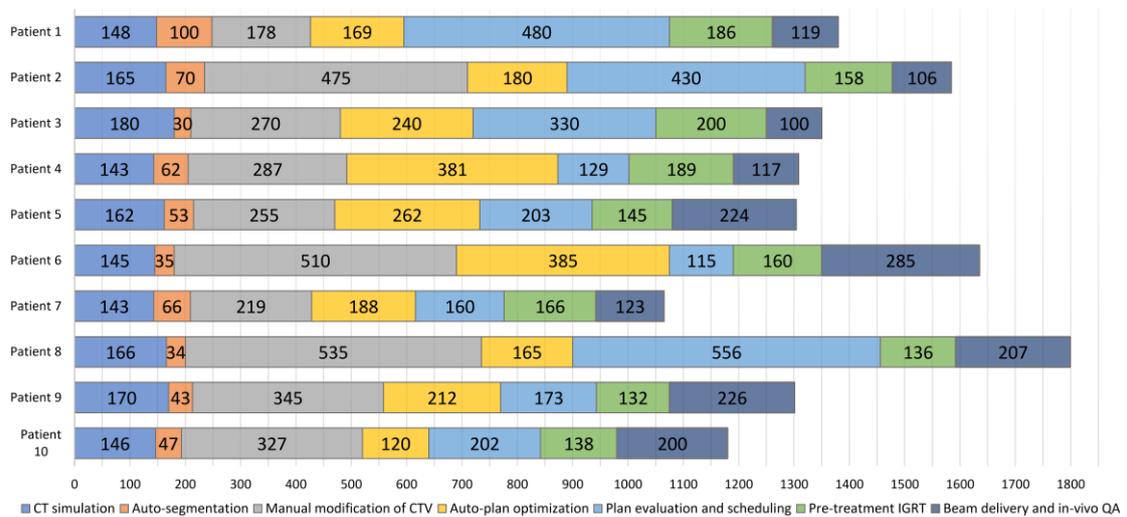

Fig. 3: Time expended (in seconds) in each phase of the All-in-One workflow

## 4. Discussion

**4.1 Benefits from full-process automation of radiotherapy**

With the help of AI implementation, the All-in-One solution has significantly improved the efficiency of the clinical RT workflow. First, the All-in-One solution significantly reduces the timescale of initiating a new RT treatment from days to minutes, sparing patients from repeated appointments, anxious waiting, and shuttling between multimodality facilities, and considerably improving their medical experience. Especially for the patients who are in emergency conditions such as obstructive, hemorrhage, and bone metastasis pain, though a green channel for planning and treatment is usually provided and they may not have to wait long, the All-in-One solution helps to reduce their activity levels and start the treatment as soon as possible.

Second, the one-stop workflow mitigates the systematic uncertainties between simulation and treatment. On the one hand, for subsequent deliveries, the effect of systematic changes in patient anatomy, such as tumor progression and weight alteration after the time of simulation, is minimized due to speeding up of treatment preparation. On the other hand, the one-stop solution removes the systematic errors between different facilities, such as mechanical calibration of the room laser and correction for couch sag. This is the reason why the in vivo dosimetry of the All-in-One solution demonstrates excellent treatment accuracy, as it is sensitive to both patient-related variations and machine-related errors [17].

Additionally, the All-in-One strategy offers an alternative to the clinical practice of online adaptive radiotherapy (ART). For the currently available ART workflows that integrate either kV CBCT [18] or MRI [19], a pretreatment plan based on CT simulation outside the treatment room is always required as a reference standard. Although a synthetic CT necessary for dose calculation can be obtained from CBCT or MRI images by means of deep learning [20-21], the feasibility of CBCT-only or MRI-only treatment planning is still under exploration [22-23]. In this regard, diagnostic-quality CT is indispensable for primary planning in current practice due to its superior CT-to-density accuracy. In this study, when the target coverage and/or OAR sparing recalculated on the daily CTs exceed clinical dose tolerances, the All-in-One solution allows on-couch full replanning for robust adaption with acceptable session time and sufficient treatment accuracy, which is particularly suitable for relatively large anatomic changes.

### 4.2 Limitations of this study

The All-in-One strategy is designed as automatically as possible to save time and to minimize the chance of errors, but the critical phases need full human interventions and require the presence of dedicated staff, including radiation oncologists and medical physicists, throughout the treatment session, because it is essential for quality control of the online workflow to secure the safety and efficacy of radiotherapy, considering the potential pitfalls of AI due to its lack of transparency. The costs of resources and personnel limit its wide-scale practice. The situation is expected to be improved in the future by redistribution of resources and staff, as well as applications of telehealth and AI assistance.

### 4.3 Future work

This study was based on the clinical data of the enrolled patients with rectal cancer. The All-in-One solution can be implemented for other treatment sites as long as appropriate segmentation models and protocols of contouring and planning are prepared. We are now expanding the clinical cohorts to patients with breast cancer. For more complex cases with multiple target delineations and potential dose conflicts (such as lung and nasopharynx), manual adjustment could be time-consuming, and the entire session may be overrun. In that case, the All-in-One solution may not be the best choice for on-couch initial treatment, but is still applicable for adaptive replanning by propagating contours and objectives from the original plan.

**5. Conclusion**

A promising online radiotherapy workflow that enables the full-process automation based on a hybrid CT-linac is proposed. The clinical results demonstrate that an accurate and efficient treatment can be delivered to the patient in a one-stop workflow. The All-in-One solution has refined the workflow effectiveness and showed potential in promoting the clinical application of online adaptive radiotherapy.

**Reference:**

1. Nelms BE, Robinson G, Markham J, et al. Variation in external beam treatment plan quality: An inter-institutional study of planners and planning systems. Pract Radiat Oncol. 2012;2(4):296-305. doi:10.1016/j.prro.2011.11.012

2. Kishan AU, Cui J, Wang PC, Daly ME, Purdy JA, Chen AM. Quantification of gross tumour volume changes between simulation and first day of radiotherapy for patients with locally advanced malignancies of the lung and head/neck. J Med Imaging Radiat Oncol. 2014;58(5):618-624. doi:10.1111/1754-9485.12196

3. González Ferreira JA, Jaén Olasolo J, Azinovic I, Jeremic B. Effect of radiotherapy delay in overall treatment time on local control and survival in head and neck cancer: Review of the literature. Rep Pract Oncol Radiother. 2015;20(5):328-339. doi:10.1016/j.rpor.2015.05.010

4. Dawson LA, Sharpe MB. Image-guided radiotherapy: rationale, benefits, and limitations. Lancet Oncol. 2006;7(10):848-858. doi:10.1016/S1470-2045(06)70904-4

5. Liu F, Ahunbay E, Lawton C, Li XA. Assessment and management of interfractional variations in daily diagnostic-quality-CT guided prostate-bed irradiation after prostatectomy. Med Phys. 2014;41(3):031710. doi:10.1118/1.4866222

6. Alderliesten T, den Hollander S, Yang TJ, et al. Dosimetric impact of post-operative seroma reduction during radiotherapy after breast-conserving surgery. Radiother Oncol. 2011;100(2):265-270. doi:10.1016/j.radonc.2011.03.008

7. Francolini G, Desideri I, Stocchi G, et al. Artificial Intelligence in radiotherapy: state of the art and future directions. Med Oncol. 2020;37(6):50. Published 2020 Apr 22. doi:10.1007/s12032-020-01374-w

8. Cardenas CE, Yang J, Anderson BM, Court LE, Brock KB. Advances in Auto-Segmentation. Semin Radiat Oncol. 2019;29(3):185-197. doi:10.1016/j.semradonc.2019.02.001

9. Meyer P, Biston MC, Khamphan C, et al. Automation in radiotherapy treatment planning: Examples of use in clinical practice and future trends for a complete

automated workflow. Cancer Radiother. 2021;25(6-7):617-622. doi:10.1016/j.canrad.2021.06.006

10. Men K, Dai J, Li Y. Automatic segmentation of the clinical target volume and organs at risk in the planning CT for rectal cancer using deep dilated convolutional neural networks. Med Phys. 2017;44(12):6377-6389. doi:10.1002/mp.12602

11. Ge Y, Wu QJ. Knowledge-based planning for intensity-modulated radiation therapy: A review of data-driven approaches. Med Phys. 2019;46(6):2760-2775. doi:10.1002/mp.13526

12. Xia X, Wang J, Li Y, et al. An Artificial Intelligence-Based Full-Process Solution for Radiotherapy: A Proof of Concept Study on Rectal Cancer. Front Oncol. 2021;10:616721. Published 2021 Feb 3. doi:10.3389/fonc.2020.616721

13. Yu L, Zhao J, Zhang Z, Wang J, Hu W. Commissioning of and preliminary experience with a new fully integrated computed tomography linac. J Appl Clin Med Phys. 2021;22(7):208-223. doi:10.1002/acm2.13313

14. Han M, Yao G, Zhang W, et al. Segmentation of CT thoracic organs by multi-resolution VB-nets. SegTHOR@ISBI 2019.

15. Feng B, Yu L, Mo E, et al. Evaluation of Daily CT for EPID-Based Transit In Vivo Dosimetry. Front Oncol. 2021;11:782263. Published 2021 Nov 2. doi:10.3389/fonc.2021.782263

16. Kehwar TS. Analytical approach to estimate normal tissue complication probability using best fit of normal tissue tolerance doses into the NTCP equation

of the linear quadratic model. J Cancer Res Ther. 2005;1(3):168-179. doi:10.4103/0973-1482.19597

17. Celi S, Costa E, Wessels C, Mazal A, Fourquet A, Francois P. EPID based in vivo dosimetry system: clinical experience and results. J Appl Clin Med Phys. 2016;17(3):262-276. Published 2016 May 8. doi:10.1120/jacmp.v17i3.6070

18. Archambault Y, Boylan C, Bullock D, et al. Making on-line adaptive radiotherapy possible using artificial intelligence and machine learning for efficient daily re-planning. Med Phys Intl J. 2020;8(2)

19. Winkel D, Bol GH, Kroon PS, et al. Adaptive radiotherapy: The Elekta Unity MR-linac concept. Clin Transl Radiat Oncol. 2019;18:54-59. Published 2019 Apr 2. doi:10.1016/j.ctro.2019.04.001

20. Gao L, Xie K, Wu X, et al. Generating synthetic CT from low-dose cone-beam CT by using generative adversarial networks for adaptive radiotherapy. Radiat Oncol. 2021;16(1):202. Published 2021 Oct 14. doi:10.1186/s13014-021-01928-w

21. Boulanger M, Nunes JC, Chourak H, et al. Deep learning methods to generate synthetic CT from MRI in radiotherapy: A literature review. Phys Med. 2021;89:265-281. doi:10.1016/j.ejmp.2021.07.027

22. Tyagi N, Fontenla S, Zelefsky M, et al. Clinical workflow for MR-only simulation and planning in prostate. Radiat Oncol. 2017;12(1):119. Published 2017 Jul 17. doi:10.1186/s13014-017-0854-4

23. Lei Y, Harms J, Wang T, et al. MRI-only based synthetic CT generation using dense cycle consistent generative adversarial networks. Med Phys. 2019;46(8):3565-3581. doi:10.1002/mp.13617